\documentclass[10pt]{iopart}

\usepackage{graphicx}

\begin{document}

\title{Different Realizations of Cooper-Frye Sampling with Conservation Laws}
\author{C. Schwarz$^{1,2}$, D. Oliinychenko$^{1,5}$, L.-G. Pang$^{1,3,4}$,  S. Ryu$^{1}$ and H. Petersen$^{1,2,6}$}

\address{$^1$ Frankfurt Institute for Advanced Studies, Ruth-Moufang-Stra{\ss}e 1, 60438 Frankfurt am Main, Germany}
\address{$^2$ Institut f\"ur Theoretische Physik, Goethe-Universit\"at, Max-von-Laue-Stra{\ss}e 1, 60438 Frankfurt am Main, Germany}
\address{$^3$ Lawrence Berkeley National Laboratory, Berkeley, CA 94720, USA}
\address{$^4$ Department of Physics, University of California, Berkeley, CA 94720, USA}
\address{$^5$ Bogolyubov Institute for Theoretical Physics, Kiev 03680, Ukraine}
\address{$^6$ GSI Helmholtzzentrum f\"ur Schwerionenforschung GmbH, Planckstra{\ss}e 1, 64291 Darmstadt, Germany}
\ead{petersen@fias.uni-frankfurt.de}


\begin{abstract}
  Approaches based on viscous hydrodynamics for the hot and dense stage and
  hadronic transport for the final dilute rescattering stage are successfully
  applied to the dynamic description of heavy ion reactions at high beam energies.
  One crucial step in such hybrid approaches is the so called particlization, the
  transition between the hydrodynamic description to microscopic degrees of
  freedom. For this purpose, individual particles are sampled on the Cooper-Frye
  hypersurface. In this work, 4 different realizations of sampling algorithms are
  compared, where three of them incorporate global conservation laws of quantum
  numbers in each event. The algorithms are compared within two types of scenarios:
  simple ``box'' hypersurface consisting of only one static cell and a typical
  particlization hypersurface for Au+Au collisions at $\sqrt{s_\mathrm{NN}} = 200$ GeV.
  For all algorithms the mean multiplicities (or particle spectra)
  remain unaffected by global conservation laws in the case of large volumes.
  In contrast, the fluctuations of the particle numbers are affected considerably.
  The fluctuations of the newly developed SPREW algorithm based on exponential weights and the recently suggested SER algorithm based on ensemble rejection
  are smaller than without conservation laws and agree with the expectation from the
  canonical ensemble. The previously applied mode sampling algorithm
  produces dramatically larger fluctuations, than it is expected in the corresponding microcanonical ensemble,
  and therefore should be avoided in fluctuation studies. This study might be of interest for
  investigations of particle fluctuations and correlations, e.g. the suggested signatures for a phase transition or a critical endpoint, in hybrid approaches
  that are affected by global conservation laws.
\end{abstract}


\ioptwocol
\maketitle

\section{Introduction}
\label{intro}

Bulk observables in heavy ion reactions at high beam energies carried out at
the Relativistic Heavy Ion Collider (RHIC) and the Large Hadron Collider (LHC)
are successfully described by hybrid approaches (see reviews in
\cite{Hirano:2012kj, Petersen:2014yqa} and references therein). Event-by-event
calculations based on relativistic viscous fluid dynamics starting from
fluctuating initial conditions and coupled to hadronic transport approaches for
the late dilute stages are the current state-of-the-art for a realistic dynamic
description of these heavy ion reactions
\cite{Werner:2010aa,Shen:2014vra,Ryu:2015vwa}. Nowadays, such calculations are
performed in a multi-parameter space and by detailed comparisons to a plethora
of observables quantitative constraints on the properties of hot and dense QCD
matter are extracted by Bayesian techniques
\cite{Pratt:2015zsa,Bernhard:2015hxa,Bernhard:2016tnd}.

There are two crucial interfaces in such combined microscopic and macroscopic approaches: The
construction of the initial state for hydrodynamics accompanied by rapid
thermalization \cite{Oliinychenko:2015lva} and the particlization at the
transition from the hydrodynamic description to individual particles. The first
topic is under heavy investigation especially also in the context of the
collective effects observed in small systems at RHIC and LHC
\cite{Keegan:2016cpi,Schenke:2016ksl}. In this work, we concentrate on the
second transition, where the sampling of particles on the hypersurface is
usually performed according to the Cooper-Frye formula \cite{Cooper:1974mv}.
Physics-wise the transition from hydrodynamics to the non-equilibrium transport
description should take place, when the degrees of freedom are hadrons and the
Knudsen number grows too large for fluid dynamics to be aplicable
\cite{Ahmad:2016ods}. The chemical and kinetic freeze-out is then subsequently
performed automatically in the transport approach
\cite{Bass:2000ib,Steinheimer:2017vju}.

Usually the particles are sampled on the hypersurface according to
grand-canonical distribution functions taking the flow velocity of the
corresponding cell into account. In the grand-canonical ensemble temperature,
volume and chemical potentials are fixed, but quantum numbers, energy and
momentum are only conserved on the average over many events. Therefore in
single events a discontinuity in the total energy, momentum and quantum numbers
occurs at particlization. This contradicts the philosophy behind the hybrid
approaches that particlization is just a smooth change of the formalism from
hydrodynamics to transport, but not a physical transition.  Additionally, at
the first glance it seems simply unphysical to violate conservation laws.
Nevertheless, the grand canonical sampling is justified in many cases, being a
good and computationally fast approximation. Exploring the validity region of
this approximation is one of the purposes of this article.

It is generally expected that the account of global conservation laws on an
event-by-event basis does not change averages over many events significantly. Therefore,
grand-canonical sampling should be a good approximation for bulk observables,
such as transverse momentum and rapidity spectra. In
the present manuscript it is verified that the relative difference for central Au+Au
collisions at the highest RHIC energy, where hybrid approaches are often
applied, does not exceed 2\%.  However, the error may become larger at lower
collision energies (e.g. at which a hybrid approach
\cite{Karpenko:2015xea} was recently applied), for smaller systems or for rare hadron species.
 On the other hand, in most cases
there are many particle distributions sampled per hydro run, the so called
``over-sampling'' technique to increase the statistics
\cite{Shen:2014vra,Karpenko:2015xea,Hirano:2005xf,Nonaka:2006yn,Song:2010mg,Ryu:2012at,vanderSchee:2013pia,Pang:2014pxa,Batyuk:2016qmb}.
With the oversampling factor $N_{over}$ the error decreases as $\frac{1}{\sqrt{N_{over}}}$,
therefore for bulk observables one can increase the accuracy of the grand-canonical sampling
approximation by increasing $N_{over}$.

Fluctuation and correlation observables are currently under intense
experimental investigation, because they are associated with the possible
existence of a first order phase transition between the hadron gas and the
quark gluon plasma or a critical endpoint in the QCD phase diagram \cite{Adamczyk:2013dal,Aggarwal:2010wy}. Contrary to
bulk observables, fluctuations and correlations cannot be reliably studied in a
hybrid approach with grand-canonical sampling. Spurious event-by-event
fluctuations of the energy, momentum and quantum numbers at particlization
change fluctuations of multiplicities in an uncontrollable way \cite{Steinheimer:2017dpb}.
Besides the fluctuations themselves, it may bias the selection of events according to
centrality classes.

Therefore, obeying global event-by-event conservation laws at particlization is
necessary to study multiplicity fluctuations with a hybrid approach.  To study
correlations, which typically make use of local variables, even this is not
enough.  Local conservation laws have to be ensured, as it was successfully tried for
electric charge in \cite{Bozek:2012en}.

Event-by-event conservation laws at particlization are important to
study certain observables using hybrid approaches. This was the motivation in
the original event-by-event hybrid approach based on the UrQMD transport model
\cite{Petersen:2008dd,Huovinen:2012is}, where one initial state is propagated
through a hydrodynamic evolution and one final state is sampled for each event.
Conservation laws (except momentum) are obeyed globally in this approach via
the so-called ``mode sampling'' algorithm \cite{Huovinen:2012is}. Recently it
was observed that the multiplicity fluctuations produced by this algorithm
depend on internal details \cite{Oliinychenko:2016vkg}, therefore, in this work a new algorithm to conserve global quantum numbers is suggested.

The new SPREW (Single Particle Rejection with Exponential Weights)
algorithm is proposed based on a typical grand-canonical sampling
algorithm \cite{Pang_inprep}, enhanced by conservation laws. Exponential weights
are introduced for each quantum number to favor configurations that fulfill the
conservation laws. The conventional and the new algorithm are compared to the
mode sampling algorithm and SER (Sequential Ensemble Rejection) algorithm.
In Section \ref{sec_algorithms} all realizations of
particle sampling on the Cooper-Frye hypersurface are explained with more
emphasis on the newly developed SPREW algorithm. As a first basic test in
Section \ref{test_thermal}, all three algorithms are applied to a single static
cell and the mean multiplicities as well as their fluctuations are compared to the
thermal expectation. In Section \ref{sec_results} the different realizations
are compared for Au+Au collisions at $\sqrt{s_\mathrm{NN}} = 200$ GeV and
Section \ref{sec_conclusion} summarizes the main findings.

\section{The different sampling algorithms}
\label{sec_algorithms}

The purpose of any particlization algorithm is to produce particles given a set
of hydrodynamic variables on a predefined hypersurface: energy density
$\epsilon(x)$, pressure $p(x)$, temperature $T(x)$, chemical potentials
$\mu(x)$ and collective velocities $u^{\mu}(x)$. The hypersurface is
numerically divided into many small pieces characterized by the local normal
vectors $d\sigma_{\mu}(x)$. According to the Cooper-Frye formula, the average
number of particles with 4-momentum $p^{\mu}$ from one cell is

\begin{equation} \label{eq:CF}
  dN = g \frac{p^{\mu} d\sigma_{\mu}}{(2\pi\hbar)^3} \mathit{f}(p^{\mu}u_{\mu}) \frac{d^3p}{p^0} \,,
\end{equation}
where $\mathit{f}(p^{\mu}u_{\mu})$ is the distribution function and $g$ is the degeneracy
of a hadron species. The average number of hadrons of species $i$ from a cell is

\begin{equation} \label{eq:density}
  \overline{n}_i = \frac{g_i u^{\mu} d\sigma_{\mu}}{2\pi^2 \hbar^3} \int \mathit{f}(p^0) p^2 dp \,,
\end{equation}
which is obtained from Eq. (\ref{eq:CF}) using its Lorentz-invariance. At this
step in conventional sampling the grand-canonical ensemble is assumed,
neglecting global conservation laws. This is a challenging assumption in heavy
ion collisions, where net baryon number, electric charge, strangeness, energy or
momentum are conserved in each event.  Nevertheless, as discussed above, in many cases this assumption may
be a good approximation.

For the grand-canonical ensemble one can express probabilities of occupation
numbers $n$ via average $\overline{n}$:

\begin{equation}
  P(n) = \left\{ \begin{array}{ll}
             \frac{\overline{n}^n}{n!} e^{-\overline{n}}, & \textrm{Boltzmann stat.} \\
             \overline{n}^n (1 + \overline{n})^{-1-n},    & \textrm{Bose stat.} \\
             \overline{n}^n (1 - \overline{n})^{1-n},   & \textrm{Fermi stat.}
           \end{array} \right.
\end{equation}
Usually $P(n_i)$ for Boltzmann statistics is assumed even if Bose or Fermi
statistics is used in Eq. (\ref{eq:density}). This simplifies sampling
considerably, because the Poisson distribution is additive. This means that the total
number of particles in a cell, as well as the overall number of particles are
Poissonian variables.  From this it also follows that, e.g. the total number of
particles with charge $+1$ is also a Poissonian variable, as well as number of
particles with charge $-1$. The total charge, or baryon number therefore has a
Skellam distribution. The net strangeness, total energy and momentum are not fixed in the conventional sampling procedure, but distributed over a range of values as it has been discussed in \cite{Huovinen:2012is}. Instead they should have
values determined by the hypersurface ($X$ denotes the overall value of BSQ and $x_i$ are BSQ numbers of each
particle, where BSQ stands for net baryon, strangeness and electric charge):

\begin{equation}
  X_{tot}  = \int_{\sigma} n_i x_i \, u^{\mu} d\sigma_{\mu} \,,
\end{equation}
where the integral runs over the hypersurface. The total energy and momentum should be fixed to

\begin{equation}
 P_{tot}^{\mu} = (E_{tot}, \vec{p}_{tot})^{\mu}  = \int_{\sigma}  T^{\mu\nu} d\sigma_{\nu} \,,
\end{equation}
where the energy-momentum tensor $T^{\mu\nu}$ is provided by the hydrodynamics.
The distribution to be sampled is therefore

\begin{eqnarray} \label{eq:distribution}
 w \sim \prod_{\mathrm{particles}} g_j \frac{p_j^{\mu} d\sigma_{\mu} \, }{p^0_j} \mathit{f}(p_j^{\mu}u_{\mu}) d^3p_j
         \times  \nonumber \\
        \times \delta^{(4)}\left(\sum_{j} P^{\mu}_j - P_{tot}^{\mu}\right) \delta^{BSQ} \left( \sum_{j} x_j - X_{tot} \right)
\end{eqnarray}
Notice that in this distribution the momenta of the particles are not independent,
which makes the distribution (\ref{eq:distribution}) extremely hard to sample
exactly. Therefore the mode sampling algorithm and the newly developed
SPREW algorithm attempt to sample it approximately, fulfilling the
constraints given by the $\delta$-functions. In the following the algorithms
themselves are described.

\subsection{Conventional sampling}

The assumptions of the conventional sampling algorithm
are described above. It is realized in the following way:

\begin{enumerate}
  \item Average multiplicities for every hadron species in a cell
        are computed according to Eq. (\ref{eq:density}) and summed up to the total
        average multiplicity in a cell $\overline{n}$.
  \item The total number of particles in a cell is sampled $n \sim Poisson(\overline{n})$.
  \item For each particle the type is selected based on the probability $\frac{n_i}{n}$
        (an equivalent alternative to get the multiplicities of hadron species is
        to sample a multinomial distribution)
  \item The momentum is generated from the thermal
        distribution and boosted to the rest frame of the cell, which results in
        the proper distribution when the correct weighting factors are included.
  \item For every cell this is repeated.
\end{enumerate}

\subsection{SPREW sampling}

The newly developed single particle rejection with exponential weights (SPREW)
algorithm ensures the global conservation of
quantum numbers in the particle sampling by introducing weights, that suppress
configurations with unwanted quantum numbers. The general procedure is as for
conventional sampling, however, the particle species sampled with probability
$\frac{n_i}{n}$ can be either accepted or rejected based on the following
scheme.

Firstly, the difference of the quantum number $X$ (representing the baryon
number, electric charge or strangeness) of the so far produced particles to the
value on the hypersurface is calculated

\begin{equation}
  \Delta X = X_{\rm{particles}} - X_{\rm{surface}} \,.
\end{equation}
For a particle $i$ with quantum numbers $x_i$, the particle is rejected with the probability $1 - e^{-|\Delta X|}$, if $\Delta X$ and $x_i$
have the same sign.  This rejection is performed for every quantum number ($B$, $S$, $Q$),
 until one hadron is accepted. In the end, the quantum numbers are not
necessarily reproduced exactly, therefore, the last few particles are adjusted
by hand. If the total number of particles is too small, this is not possible,
since it would produce too large bias.

The momenta of the particles are sampled in the same way as in the conventional
method. After a particle ensemble has been obtained, all the energies and momenta
of the individual hadrons $\vec{p}_i$ are adjusted to enforce energy and
momentum conservation.  Firstly, the momenta are centralized: $\vec{p}_i =
\vec{p}_i - \frac{1}{N_{tot}} \sum_j p_j $.  Then all the particles are boosted to
the center of mass frame (denoted by the primed quantities). The momenta are then rescaled with factor
$(1+a)$ such that

\begin{equation}
 \sum_i \sqrt{(1+a)^2 \vert{\vec{p}\,'_n}\vert^2 + m_n^2} = E'_{\rm{hypersurface}}
\end{equation}
and then boosted back. This enforces energy and momentum conservation and
ensures that the on-shell condition $E^2 = p^2 + m^2$ is met for every particle.
Given that the typical value of $a$ is very small ($|a| \sim 3\% $)
in the case of Au-Au collisions, the SPREW sampling
conserves energy and momentum without deformation of the momentum space
distribution even though the energy-momentum conservation is not enforced at
the point of sampling.

It is important to underline the difference between rescaling the
energy and accounting for the energy conservation in the sampling directly.
While the first does not affect the multiplicity distribution, the second
implies the transition from canonical to microcanonical sampling leading to
narrower multiplicity distributions.

\subsection{Mode sampling}

Another way of implementing global conservation laws has been applied within the
UrQMD hybrid approach by the so called mode sampling procedure. Essentially, the
sampling procedure is performed several times (different 'modes'), that are
always terminated when the total energy is used for particle production. In the
first mode only the particles containing positive strangeness are kept, while all
others are discarded, the second one produces the corresponding number of
particles with negative strangeness to ensure net strangeness conservation. The
following two modes proceed in a similar way for net baryon number conservation
and then the same procedure is pursued for charged particles that are
non-strange mesons. Last, the energy is filled up with non-strange neutral
mesons, which are mainly $\pi_0$'s. The mode sampling is described in more
detail in \cite{Petersen:2008dd,Huovinen:2012is,Oliinychenko:2016vkg}.

\subsection{SER sampling}

One more algorithm of implementing a sampling respecting global conservation
laws can be called sequential ensemble rejection (SER).  This algorithm was
already used and studied in \cite{Oliinychenko:2016vkg} under the name
of ``unbiased Becattini-Ferroni sampling''. The algorithm proceeds as follows:

\begin{enumerate}
  \item For the quantum number $X \in \{B,S,Q\}$ repeat the following.
        Sample the total number of $N_{X>0}$ and $N_{X<0}$ from a Poisson distribution.
        The multiplicities of particular hadron species with $X>0$ and $X<0$
        are then sampled from multinomial distributions.
        If $N_{X>0} - N_{X<0} \neq X_{surface} - X_{particles}$, where
        $X_{particles}$ is the total of the previously sampled particles, then start
        from the very beginning.
  \item Sample multiplicities of neutral mesons from the Poisson distributions.
  \item If the total energy of the sampled particles $E_{sampled}$ deviates too much
        from the expected energy $E_{surface}$ then start from the beginning.
        The quantitative criterion was chosen as $|E_{sampled} - E_{surface}|/E_{surface} < 0.01 $.
        This last step can be omitted, then the energy conservation is not enforced.
        Without this last step the algorithm corresponds to a global canonical ensemble, with
        the last step it approximately corresponds to a global microcanonical ensemble.
\end{enumerate}

\section{Thermal fluctuations}
\label{test_thermal}

In this Section a simple ``box'' test scenario is investigated, where the hypersurface
is just one static cell with $d\sigma_{\mu} = (V, 0, 0, 0)$. The temperature in this
box is assumed to be $T = 150$ MeV. All the chemical potentials are supposed to be $\mu_B = \mu_S = \mu_{I3} = 0$.

The multiplicity distributions should be compared to the corresponding analytical expectations.
The conventional sampling aims at producing the grand-canonical
ensemble, where multiplicity distributions are simply Poissonian. The goal of the SPREW algorithm
is to produce a canonical ensemble in global net baryon number $B$, strangeness $S$ and electric charge $Q$.
The SER algorithm and mode sampling additionally try to conserve the total energy, so they should be compared
to the proper microcanonical ensemble. Finally, SER algorithm without the last step,
which performs rejection by energy, is to be compared to the canonical ensemble in $BSQ$.

The analytical canonical and microcanonical distributions
are hard to obtain analytically in general, but they were computed in \cite{Hauer:2007ju}
for the case of large volumes. In this case the (micro-)canonical multiplicity distributions approach a Gaussian distribution with
the grand-canonical mean and a non-trivial variance, which
is always smaller or equal than for the grand-canonical ensemble. To compute the analytical expectation
Boltzmann-Maxwell distribution was used, Eqns. (28-33) from \cite{Hauer:2007ju} were applied for canonical ensemble
and extended analogously to Eqns. (47-53) from the same article to add energy conservation.

\begin{figure*}
  \centering
  \resizebox{1.0\textwidth}{!}{%
    \includegraphics{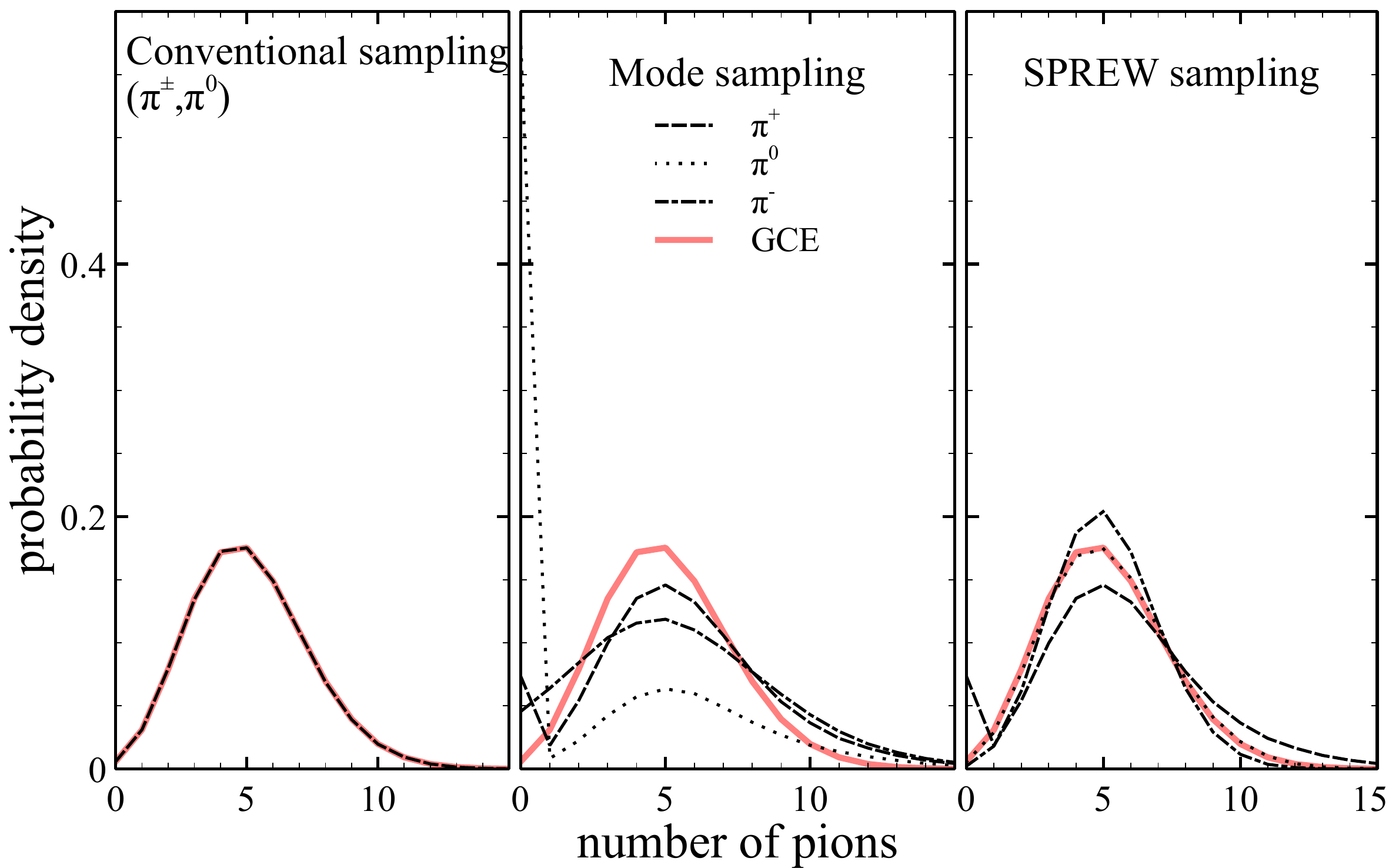}
    \includegraphics{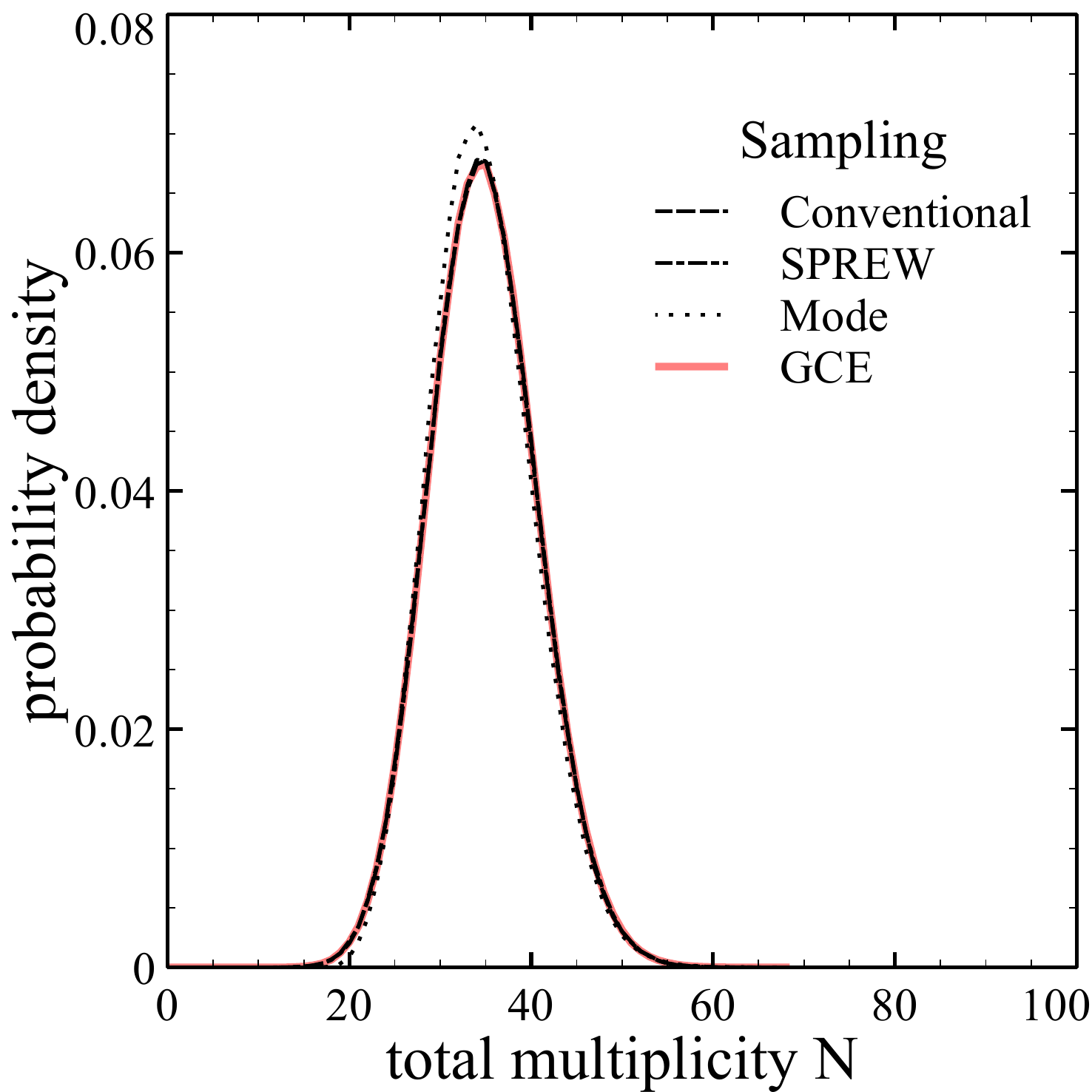}
  }
  \caption{Distribution of the number of pions (left) and the total number
           of particles (right) in a box with the length $l=5$ fm.}
  \label{fig_box_dist_5}
\end{figure*}

Let us start by investigating a small box with a length of $l=5$ fm that contains
a rather small number of particles. First, notice that theoretically the distributions of $N_{\pi^+}$
 and $N_{\pi^-}$ should be identical in any ensemble for vanishing chemical potentials.
Fig. \ref{fig_box_dist_5} (left) demonstrates that for mode sampling and SPREW these distributions
do not coincide. The deviation is the largest for the mode sampling, which does not have enough  energy left in the last modes
and fails to produce enough $\pi^-$ and especially $\pi^0$. Of course, this bias depends on the order
of the modes, as it was already noticed in \cite{Oliinychenko:2016vkg}. The SPREW algorithm
leads to a smaller deviation, which we attribute not to the algorithm itself, but rather
to the manual interventions in the final step to ensure exact conservaction of quantum numbers. We have checked that
for the SPREW sampling the results are insensitive to the order of the SPREW steps for the
different quantum numbers in contrast to the mode sampling.
The SER algorithm, which is not shown in Fig. \ref{fig_box_dist_5} produces
$N_{\pi^+}$ and $N_{\pi^-}$ distributions, which are identical to each other.

\begin{figure*}
  \centering
  \resizebox{1.0\textwidth}{!}{%
    \includegraphics{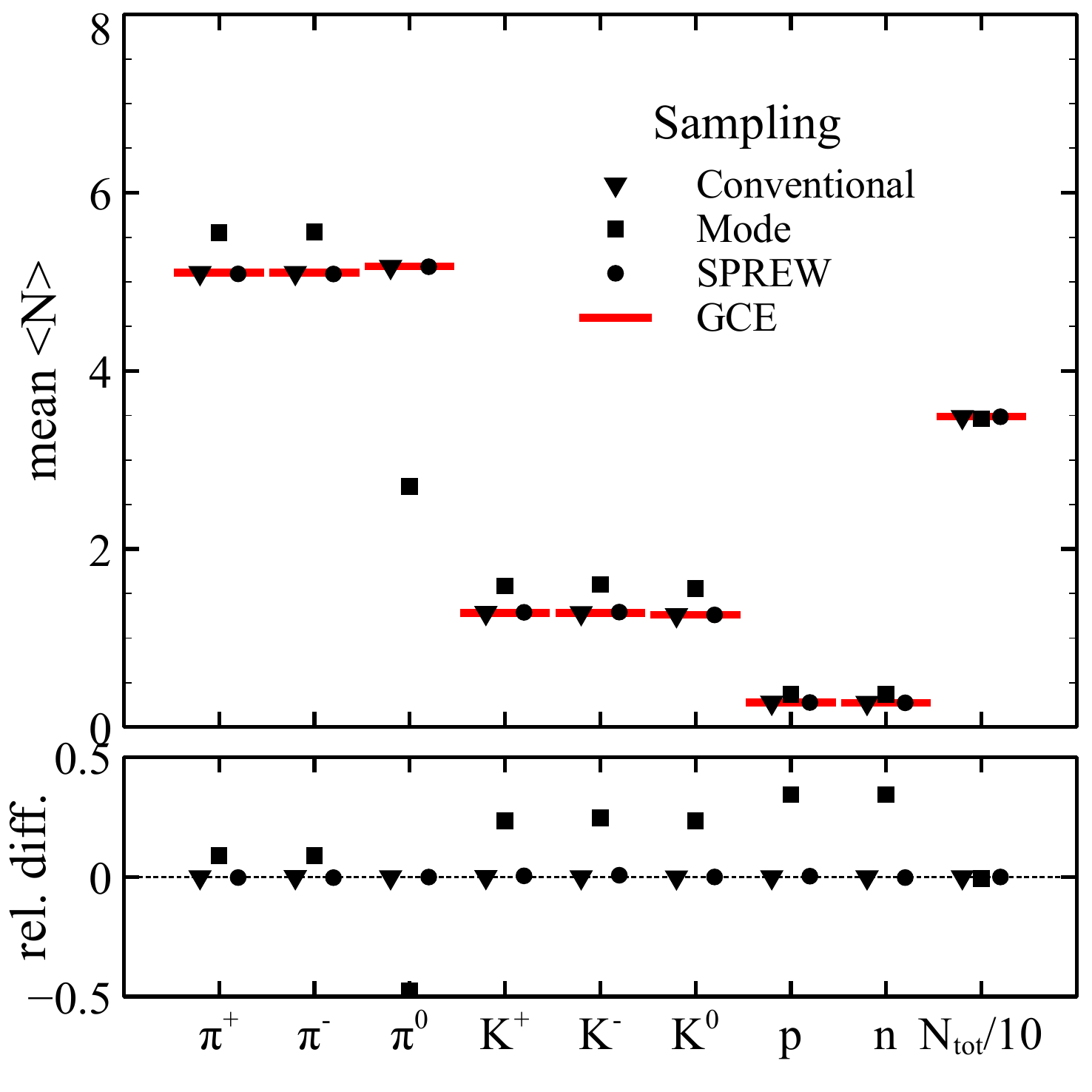}
    \includegraphics{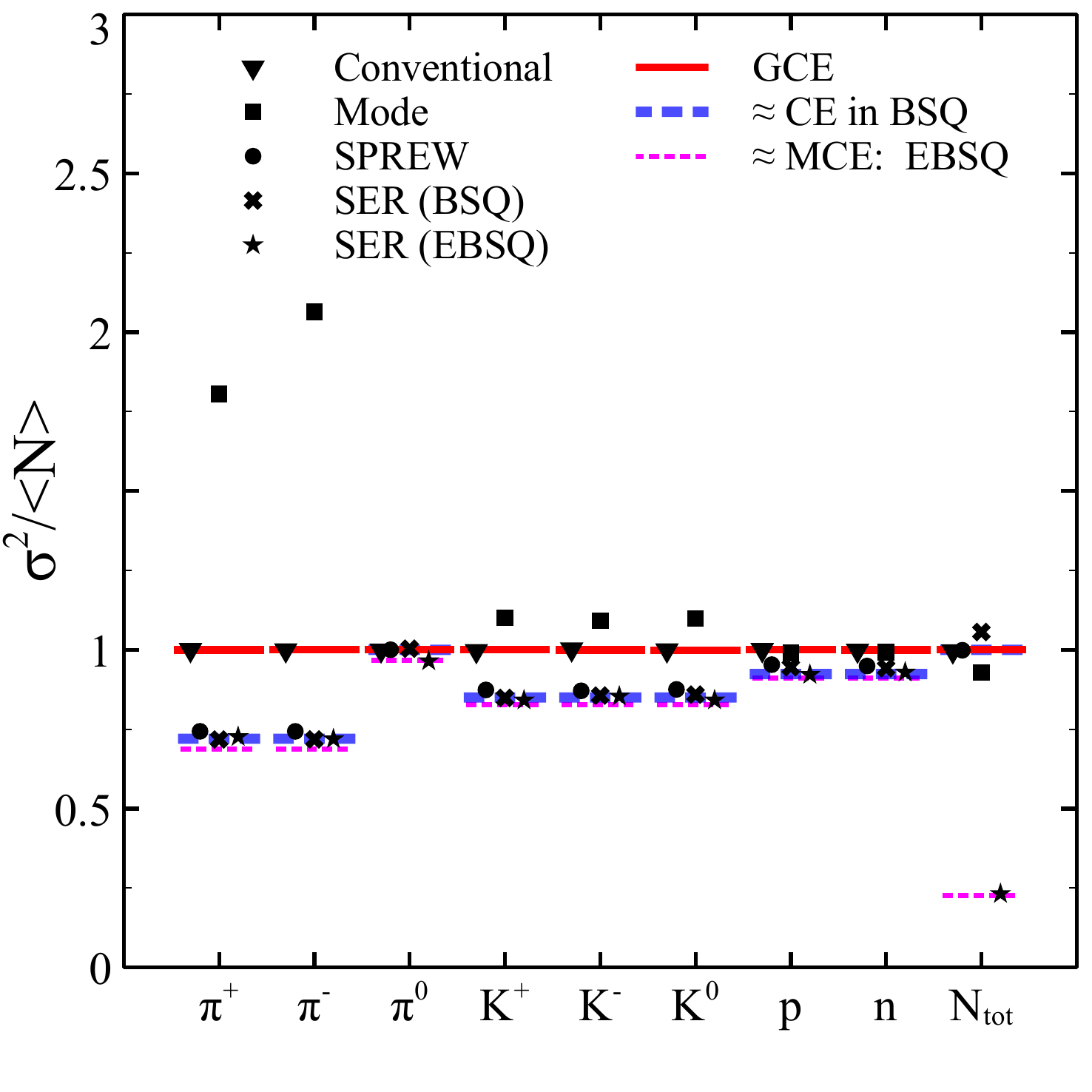}
  }
  \caption{Mean (left) and scaled variance (right) of the distributions of the
           multiplicities of different particle species in a box with the length $l=5$ fm.
           The lower panel displays the relative difference to the grand-canonical result. }
  \label{fig_box_mean_sigma_5}
\end{figure*}

Despite the small volume, the multiplicity distributions are still rather close to Gaussian distributions.
Therefore, the large volume analytical approximation should be applicable and the mean multiplicities
should be close to the grand-canonical values for all algorithms. In Fig. \ref{fig_box_mean_sigma_5} one
can see that this is indeed the case for the SPREW algorithm (the SER also fits the expectation),
but not for the mode sampling. The latter introduces a bias to the mean values in case of small particle
number. Therefore, one has to be careful when applying hybrid approaches with conservation laws
to small systems like p-p collisions.

The scaled variance $\frac{\sigma^2}{<N>} = \frac{<N^2> - <N^2>}{<N>}$ of the multiplicities is
compared in Fig. \ref{fig_box_mean_sigma_5} (right). The SPREW and SER algorithm without total energy conservation
match the approximate canonical expectation within statistical errors. The SER with total energy conservation
matches the microcanonical expectation reasonably well. By varying the margin for the energy rejection the agreement can be increased even further.
In contrast to previous algorithms, the mode sampling drastically deviates from the aimed microcanonical ensemble,
producing even wider distributions than the grand-canonical one.

\begin{figure*}
  \centering
  \resizebox{1.0\textwidth}{!}{%
    \includegraphics{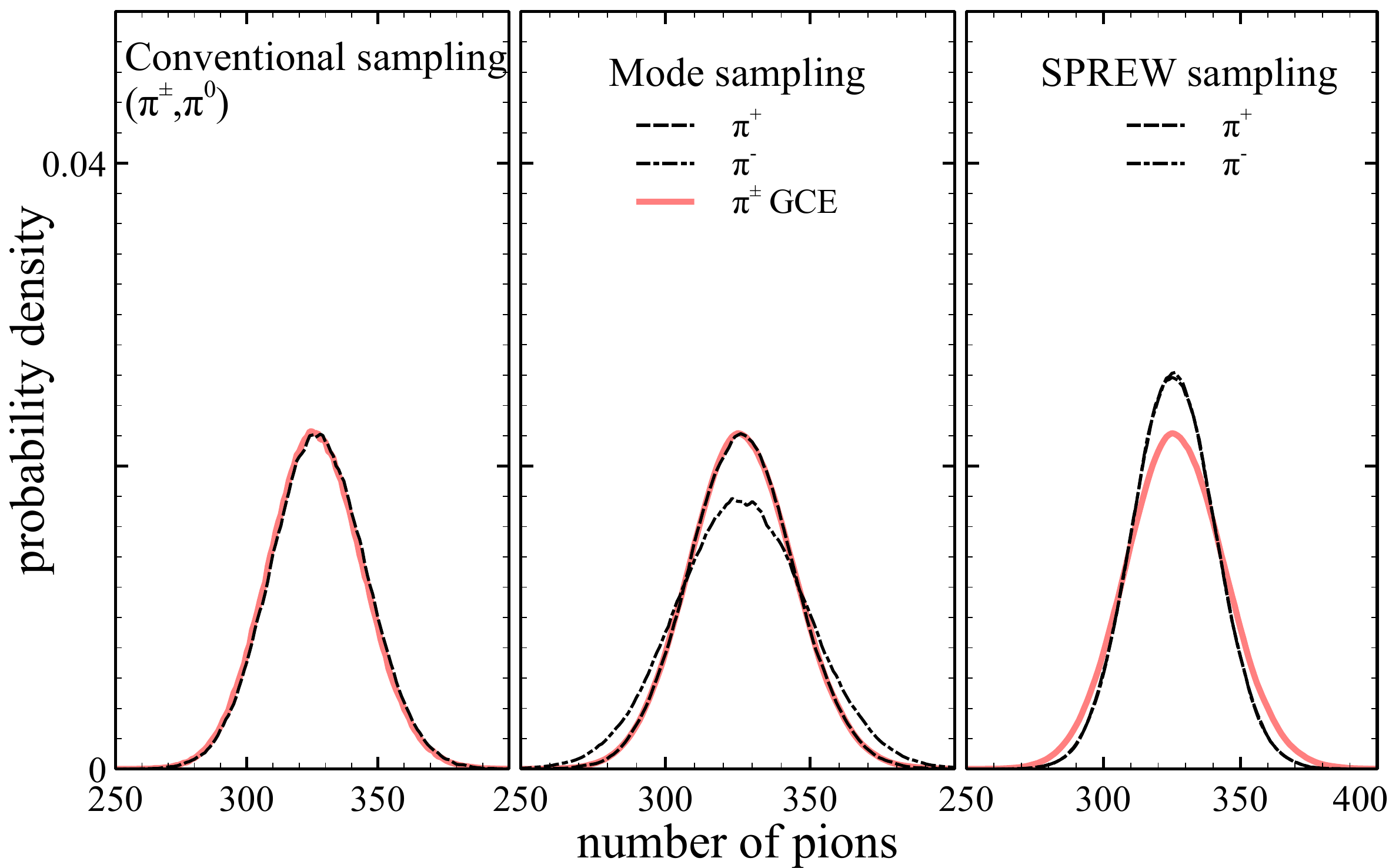}
    \includegraphics{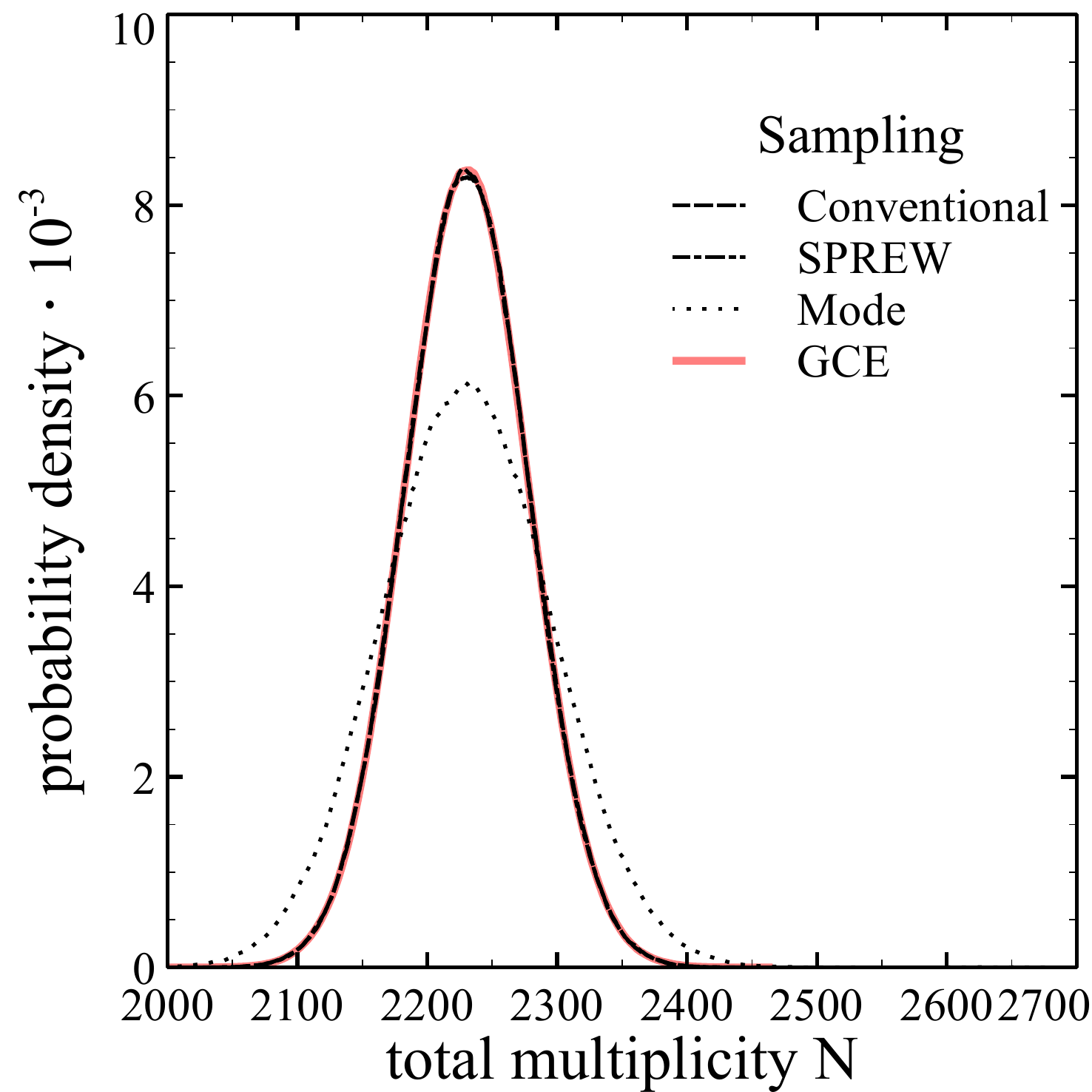}
  }
  \caption{Distribution of the number of pions (left) and the total number of particles (right) in a box with the length $l=20$ fm.}
  \label{fig_box_dist_20}
\end{figure*}

\begin{figure*}
  \centering
  \resizebox{1.0\textwidth}{!}{%
    \includegraphics{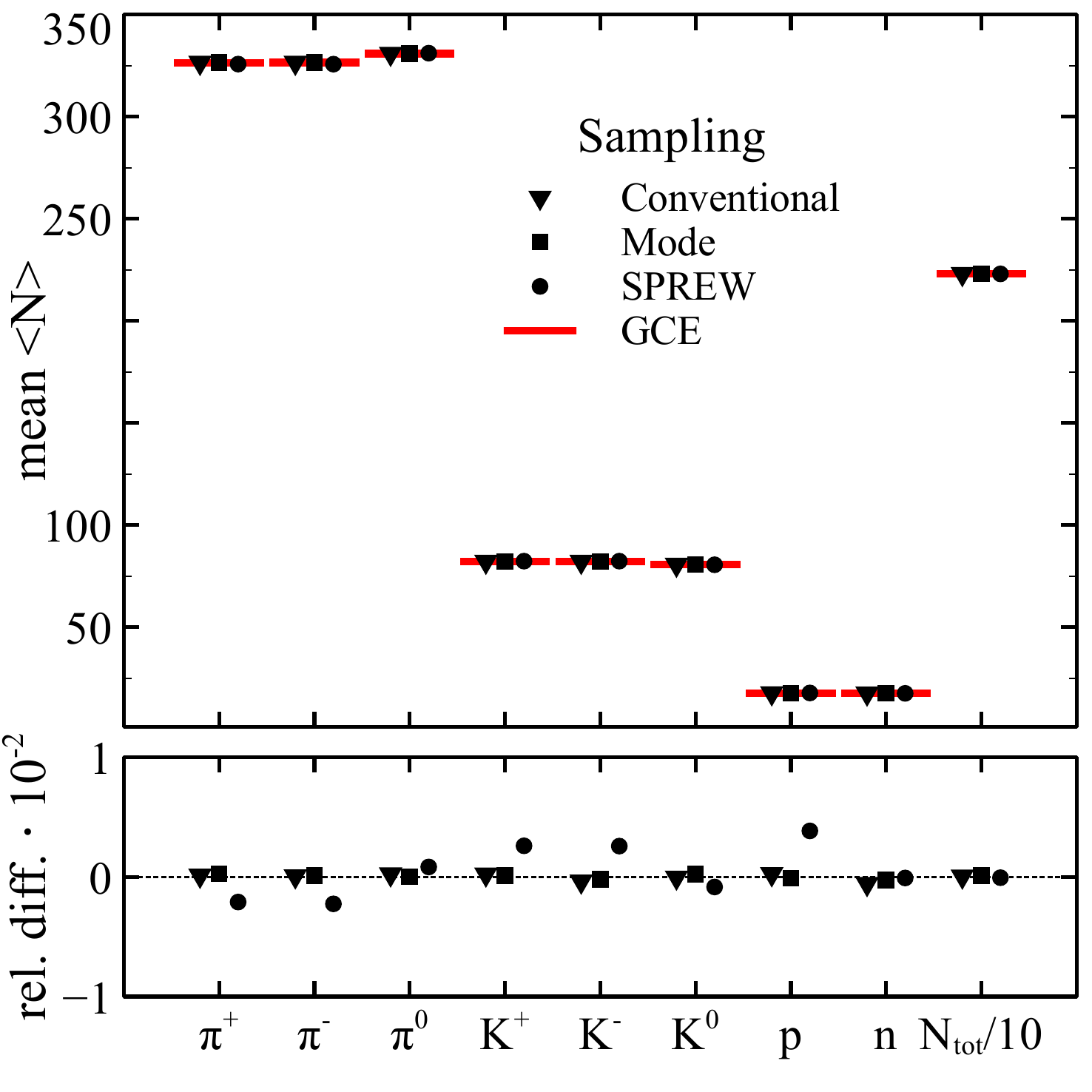}
    \includegraphics{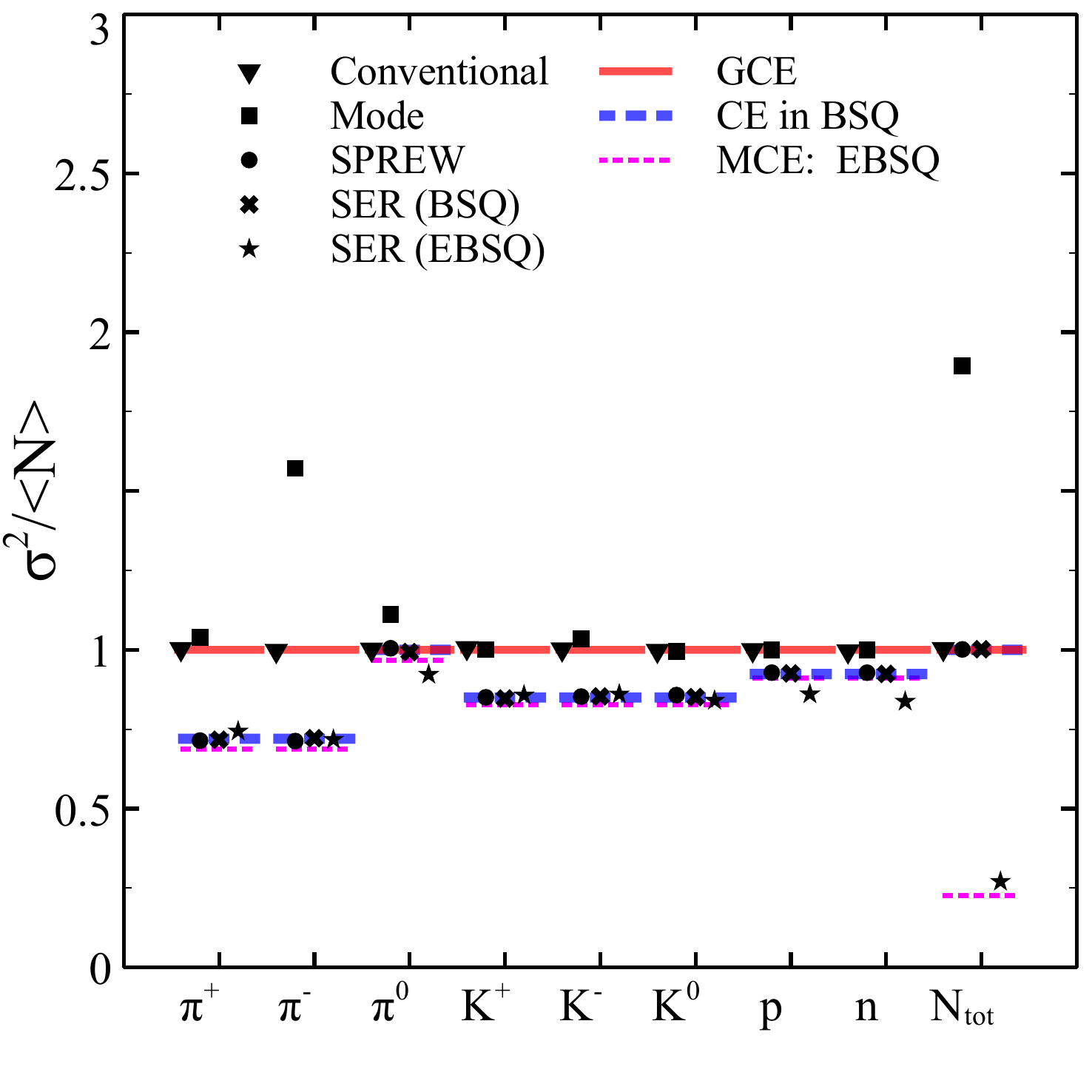}
  }
  \caption{Mean (left) and scaled variance (right) of the distributions of the multiplicities of different particle species in a box with the length $l=20$ fm. The lower panel displays the relative difference to the thermal result. }
  \label{fig_box_mean_sigma_20}
\end{figure*}

Figs. \ref{fig_box_dist_20} and \ref{fig_box_mean_sigma_20} show the same
probability distributions and quantified mean values and standard deviations in
a larger box with length $l=20$ fm that contains multiplicities similar
to the ones in a Au+Au collision at the highest RHIC energies. The multiplicity distributions
are very close to Gaussian distributions, so the large volume approximation should be accurate in this case.
The expected mean values are reproduced nicely by all algorithms within 1\% accuracy.
The fluctuations behave similarly to the smaller box, namely for the mode sampling they overshoot
the microcanonical and even the grand-canonical expectations dramatically.

In addition, we have checked that including Bose/Fermi distributions does not
change the results for the conventional sampling algorithm by employing a
completely independent implementation that has been used in \cite{Ryu:2015vwa}.

\section{Au+Au collisions at $\sqrt{s_\mathrm{NN}} = 200$ GeV}
\label{sec_results}

After testing the different realizations of sampling particles with and without
conservation laws in a box, let us now investigate the more realistic situation
of heavy ion reactions. The transition from hydrodynamics to particles is usually performed on the
Cooper-Frye hypersurface at a constant temperature.  This is a good approximation
to the hypersurface where the expansion rate exceeds the scattering rate and the
hydrodynamic evolution is not applicable anymore. To calculate the hydrodynamic
evolution and the properties on the hypersurface the CLVisc code on GPU´s
\cite{Pang:2014ipa} has been employed. Smooth Glauber initial conditions are
propagated through an ideal hydrodynamic evolution to eliminate all additional
sources of fluctuations. The hadronic rescattering has also been neglected for simplicity. The fluid
densities are converted to single particles on the Cooper-Frye hypersurface at a
freeze-out temperature of $T=137$ MeV. The parameterization with a smooth crossover
connecting lattice QCD EoS at high temperatures and a hadron resonance gas at low temperatures in chemical equilibrium (or partial chemical
equilibrium if s95p-pce is applied)~\cite{Huovinen:2009yb}, is used in the current
calculation. Particles are sampled on the hypersurface 
according to the algorithms described in Section \ref{sec_algorithms}.

\begin{figure*}
  \centering
  \resizebox{1.0\textwidth}{!}{%
    \includegraphics{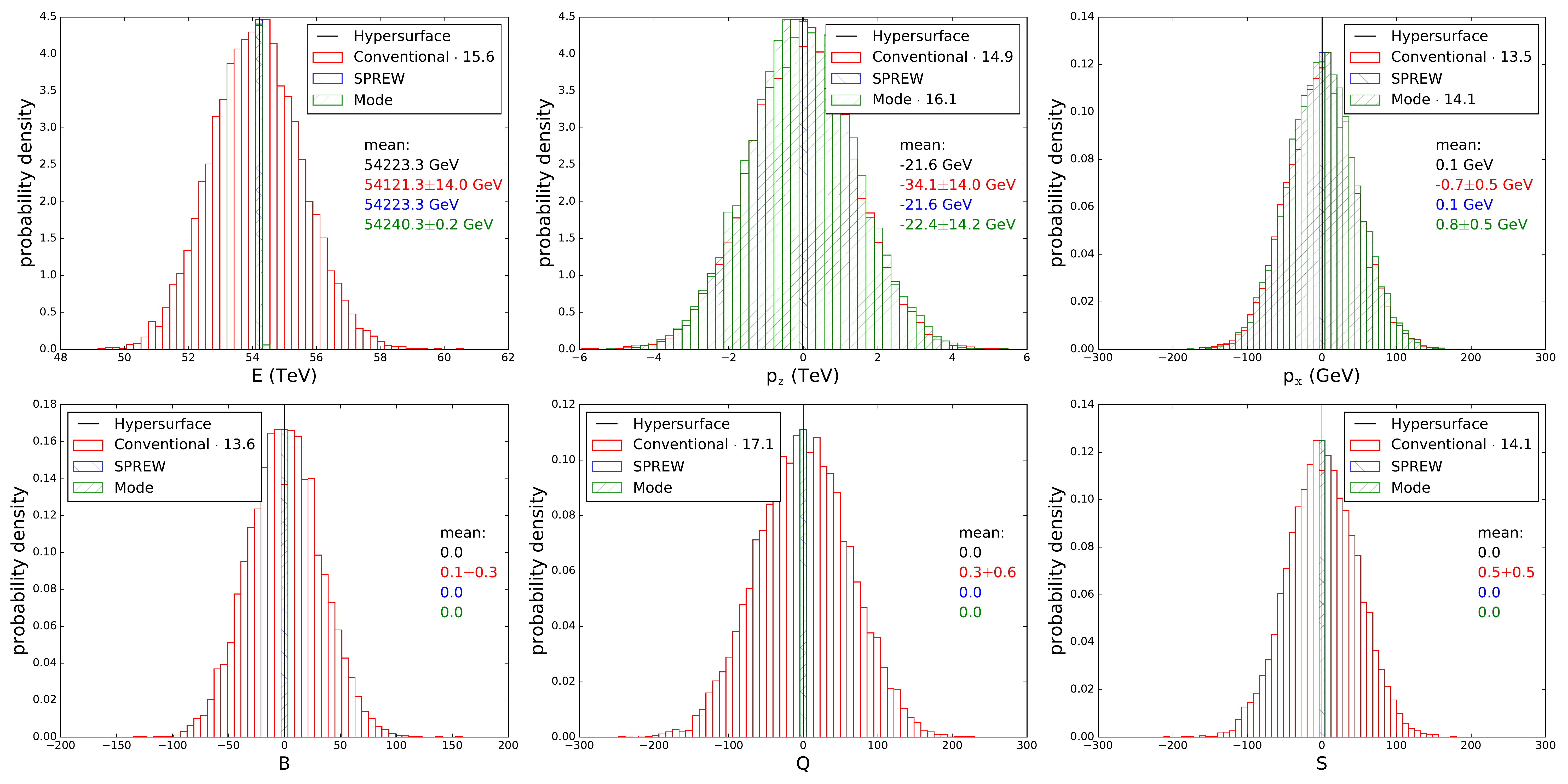}
  }
  \caption{Conserved quantities on the hypersurface in a central ($b=0$ fm)
           Au+Au collision at $\sqrt{s_\mathrm{NN}} = 200$ GeV compared to the sampled
           particles for all three algorithms. The mean values of 10.000 events are given
           as numbers in the corresponding figures.}
  \label{fig_quantum_numbers}
\end{figure*}

Let us start with a discussion of all the conserved quantum numbers. Let us
stress again, that in conventional sampling algorithms none of these is
conserved in single events, only on the average the correct result is obtained.
In Fig. \ref{fig_quantum_numbers} each of the panels contains the mean values
and their standard deviation. For the net baryon number, net strangeness and
electric charge the average values are reproduced very well. For the
energy and the momentum in z- and x- direction slight deviations can be
observed. The mode sampling conserves all quantum numbers, but the rescaling of
the momenta is not applied, which is reflected in small deviations for the total
energy on the hypersurface as well. The newly developed SPREW algorithm
nicely conserves all quantum numbers on an event-by-event basis. In the
conventional algorithm the distributions around the mean are rather wide and
allow for large fluctuations in single event particle samples. This has to be
kept in mind, when calculating more involved particle correlation and
fluctuation observables from hydrodynamic or hybrid approaches.

\begin{figure*}
  \resizebox{1.0\textwidth}{!}{%
    \includegraphics{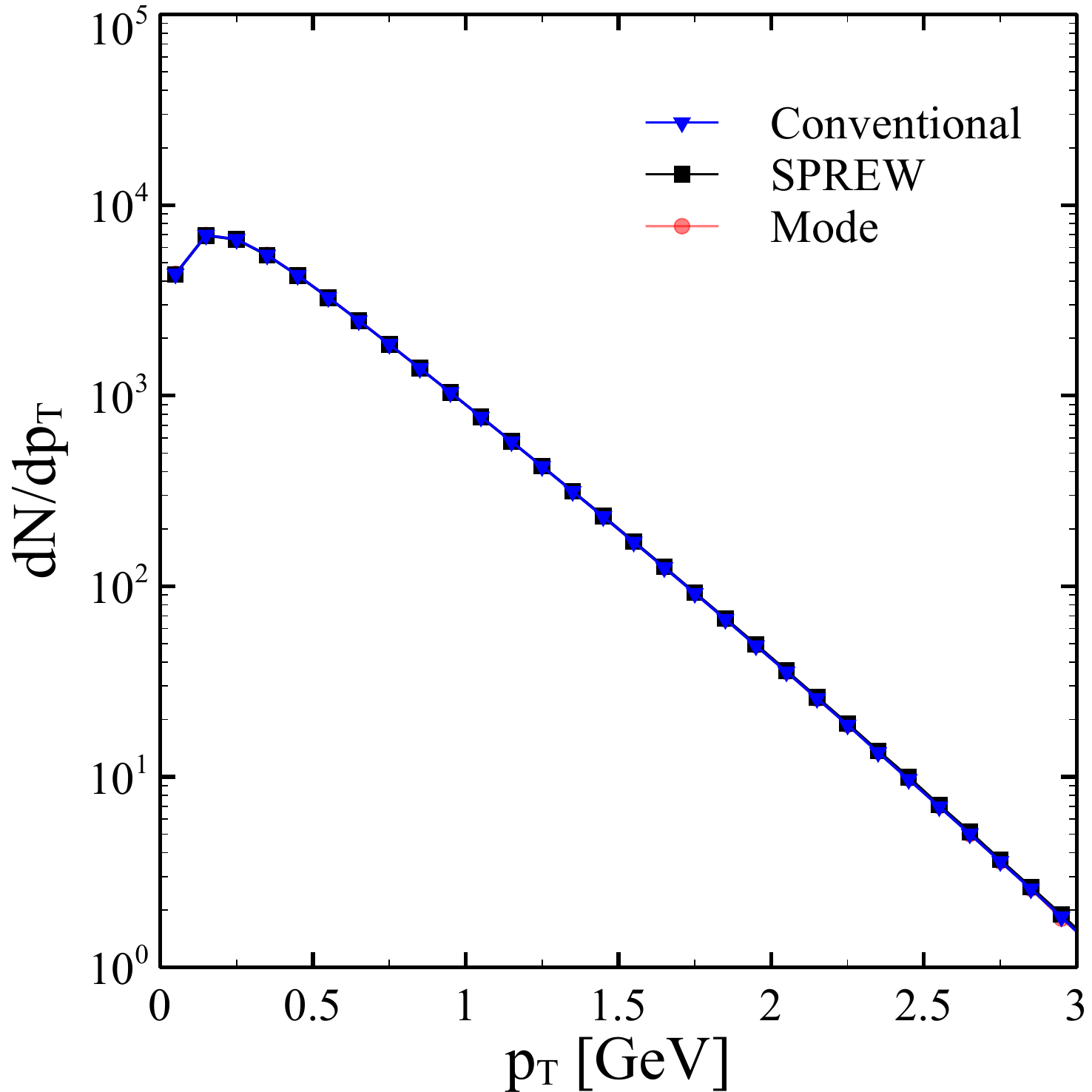}
    \includegraphics{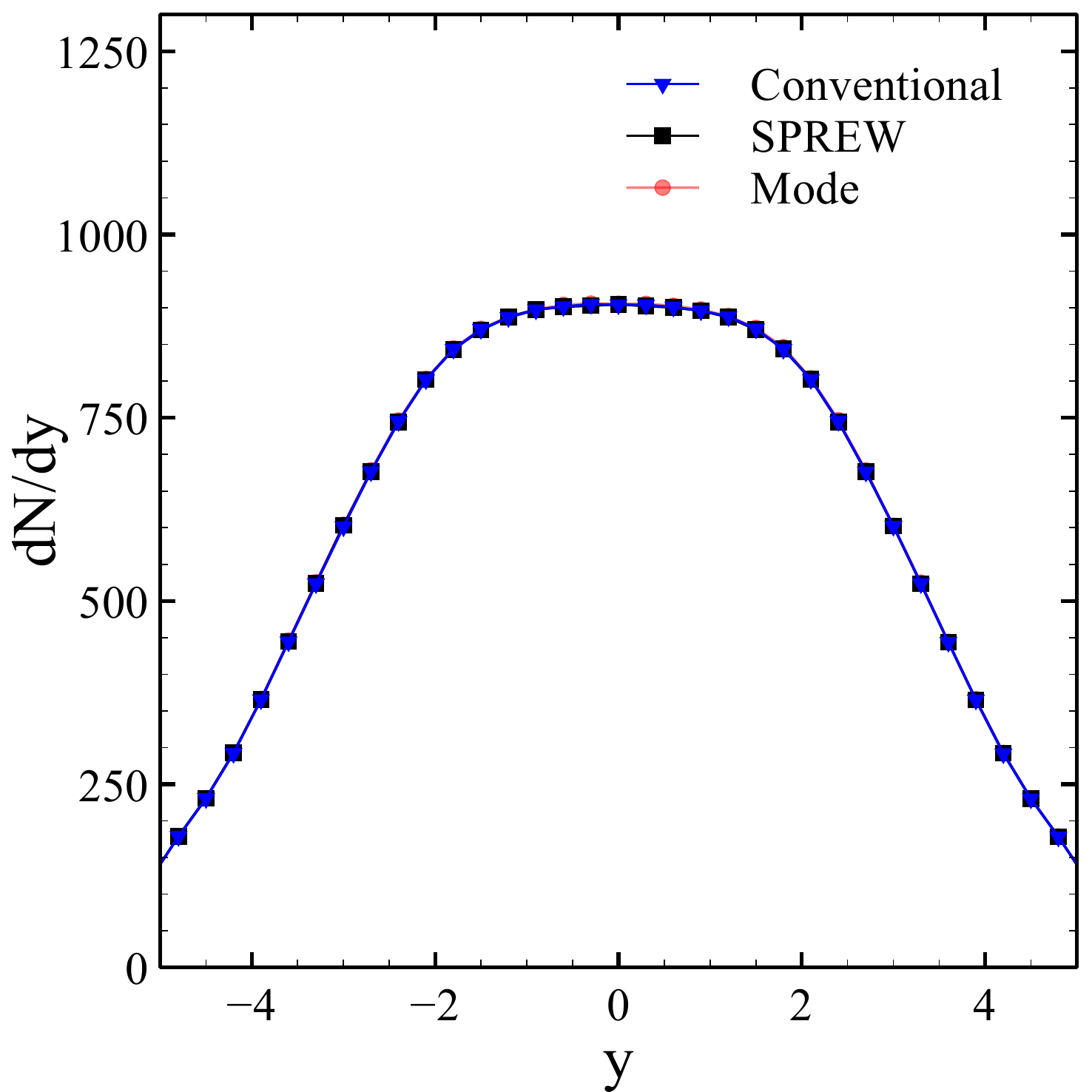}
  }

  \caption{Transverse momentum (left) and rapidity (right) distribution for all particles in a
           central ($b=0$ fm) Au+Au collision at $\sqrt{s_\mathrm{NN}} = 200$ GeV for all three algorithms.}
  \label{fig_spectra}
\end{figure*}

Here, we restrict ourselves to investigating the consequences of global
conservation laws on basic bulk observables. In Fig. \ref{fig_spectra} it can be
seen that the transverse momentum and rapidity spectra are not affected by
conservation laws. All three algorithms yield exactly the same distribution for
all particles in a central ($b=0$ fm) Au+Au collision.

\begin{figure}
  \resizebox{0.5\textwidth}{!}{%
    \includegraphics{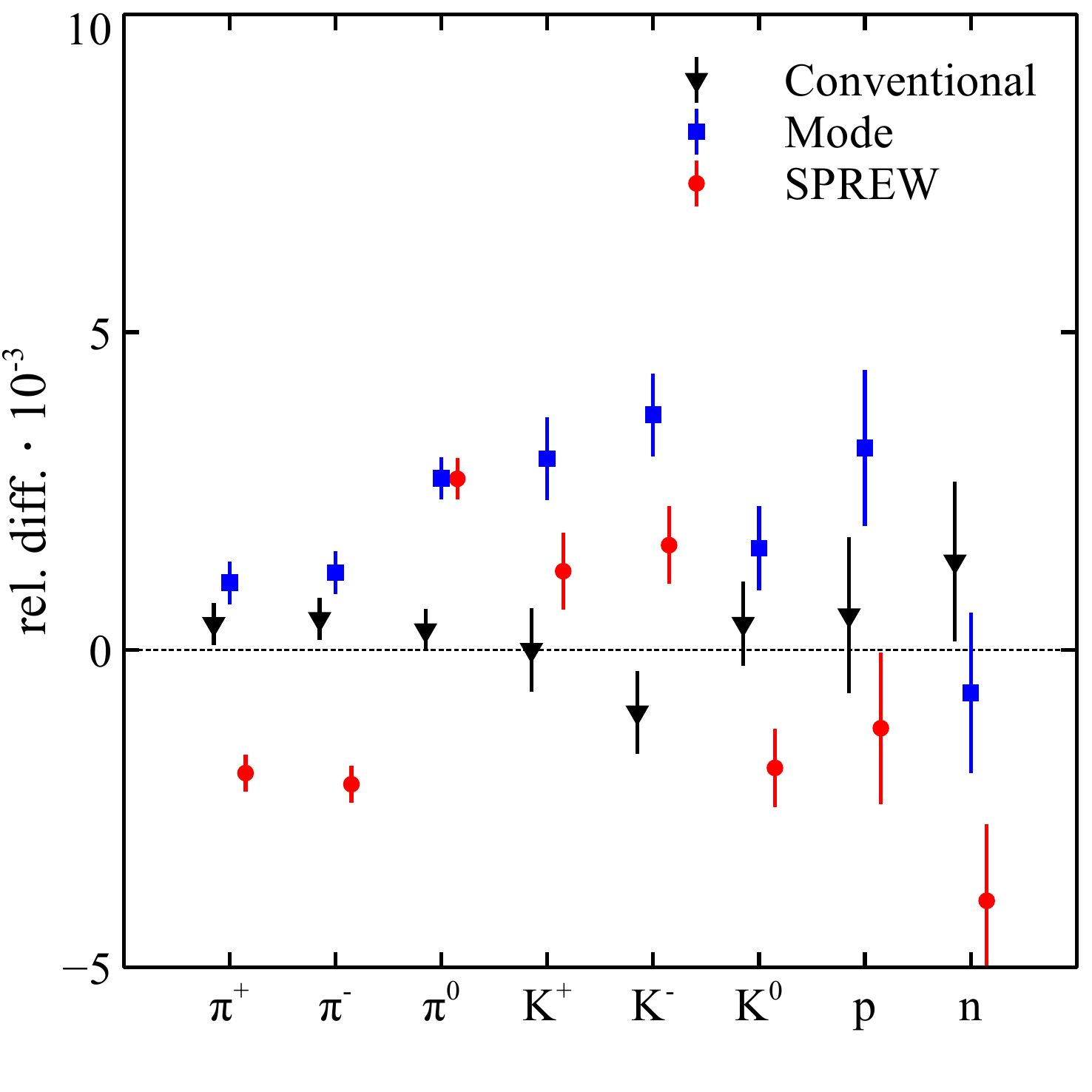}
  }
  \caption{Relative difference of the multiplicities for different most abundant
           particle species in a Au+Au collision at $\sqrt{s_\mathrm{NN}} = 200$ GeV for
           central ($b=0$ fm) for all three algorithms.}
  \label{fig_rel_mult}
\end{figure}

In Fig. \ref{fig_rel_mult} the relative multiplicities for the most abundant
particle species are shown. All of them are reproduced by all three algorithms
within a better accuracy than $0.5$ \%. This is in agreement with our finding in
the previous section, that the algorithms do not bias the particle
multiplicities, if the abundances are large enough.

\begin{figure}
  \resizebox{0.5\textwidth}{!}{%
    \includegraphics{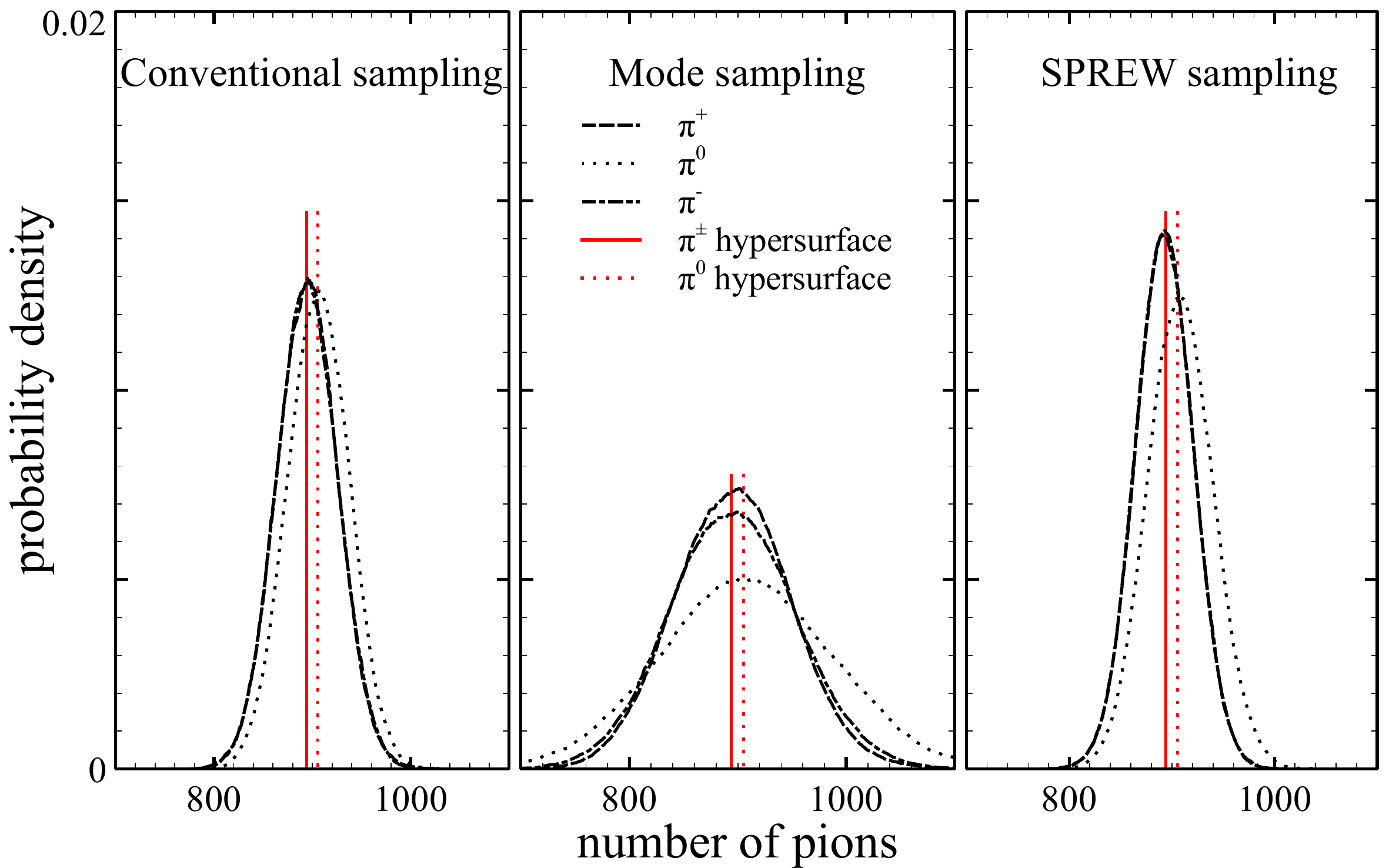}
  }
  \caption{Distributions of all three types of pions in a central ($b=0$ fm)
           Au+Au collision at $\sqrt{s_\mathrm{NN}} = 200$ GeV for all three algorithms
           compared to the integrated value on the hypersurface.}
  \label{fig_pion_dist}
\end{figure}

Fig. \ref{fig_pion_dist} shows the distributions of the different pion isospin
states in more detail, analogously to the thermal box tests in Section
\ref{test_thermal} above. First of all, a slight difference in the mean values
for $\pi^{+/-}$ and $\pi^0$ is observed that is attributed to their different
masses. Apart from the small difference in the mean values, the distributions
for the conventional sampling algorithm are very similar for the three different
types of pions. For the SPREW sampling the distributions of charged pions
are a little narrower than the original distribution, which can be attributed to
charge conservation as discussed above. The mode sampling again leads to wider
distributions and the width of the distribution is even more increased for
neutral pions. This result confirms our findings that the SPREW sampling
introduces less biases than the mode sampling, when enforcing global
conservation laws.

\begin{figure}
  \resizebox{0.5\textwidth}{!}{%
    \includegraphics{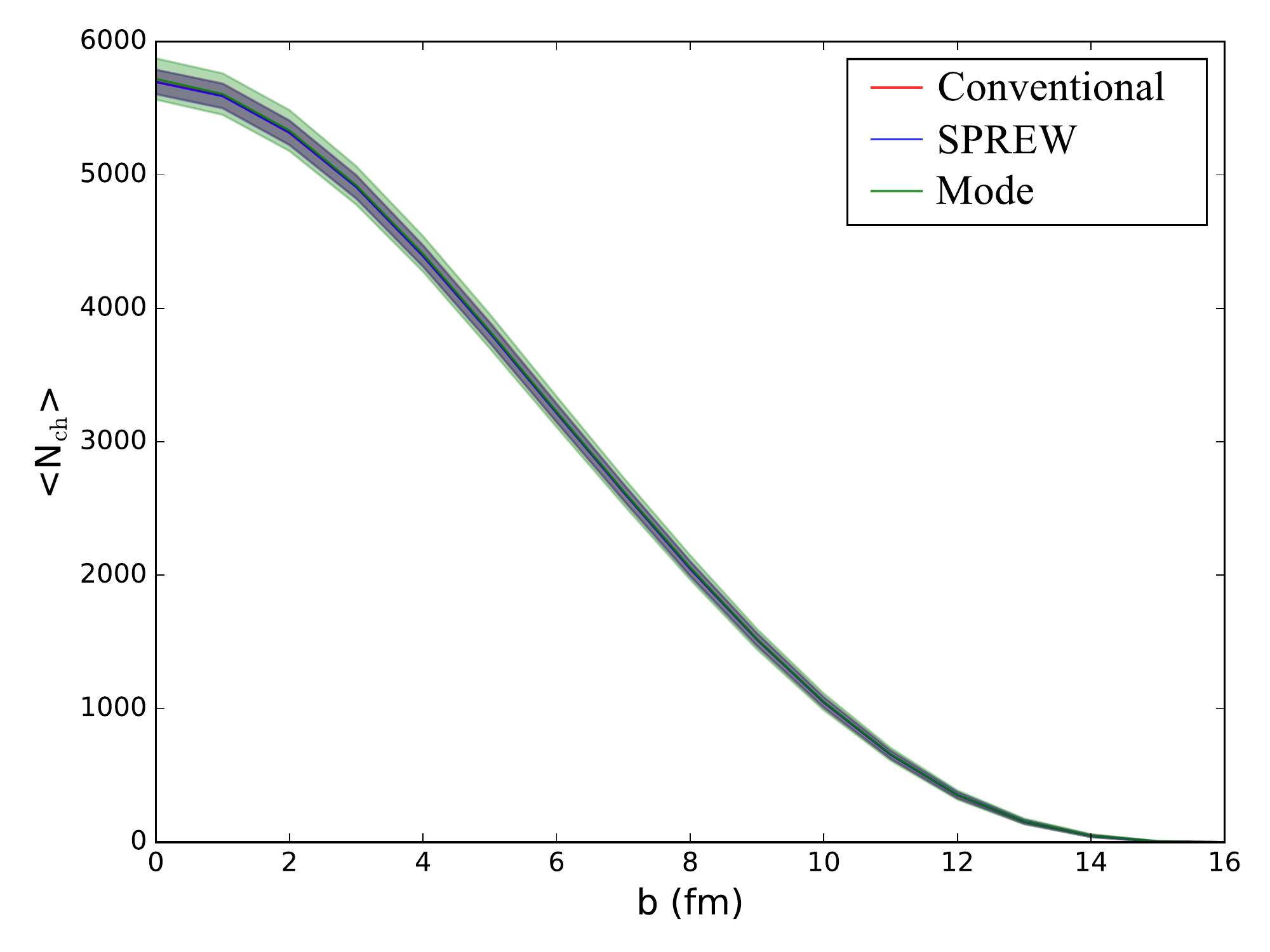}
  }

  \caption{Number of charged particles (after decays of resonances) as a function
           of the impact parameter in Au+Au collisions at $\sqrt{s_\mathrm{NN}} = 200$ GeV for
           all three algorithms. The variance of the distribution is indicated by the
           shaded area.}
 \label{fig_Nch_imp}
\end{figure}

One of the most direct consequences of different event-by-event fluctuations of
multiplicities is expected for the division of events in centrality classes. If
the fluctuations of the number of charged particles depend on the sampling
algorithm that has an effect on the selection of events for certain centrality
cuts. Especially, when extreme cuts for very central events are performed small
differences in the fluctuations can lead to different resulting event samples
for each centrality class. To quantify this effect, Fig. \ref{fig_Nch_imp} shows
the mean number of charged particles and their variance displayed by the shaded
band as a function of impact parameter for Au+Au collisions at $\sqrt{s_\mathrm{NN}} = 200$ GeV.
Resonance decays have been taken into account. The bands and
lines for the conventional and SPREW sampling overlap for all centralities,
so the SPREW sampling will not affect the selection of centrality classes.
The mode sampling on the other hand leads to wider distributions in line with
our analysis above and might affect the resulting centrality class selection.

Besides the possible effects of event-by-event global conservation laws in heavy
ion collisions, small systems might be affected much more strongly by
conservation laws. Due to the measurement of collective effects and anisotropic
flow coefficients in pp and pPb collisions at the LHC, hydrodynamic and hybrid
approaches are also applied to small systems. The effect of momentum
conservation will strongly affect the particle correlations (at least on the
2-particle level), but for this purpose local conservation laws need to be
implemented. We have nevertheless checked that global conservation laws do not
affect nor introduce structures in the $\Delta \eta-\Delta \phi$ distribution in
proton-proton collisions corresponding to a beam energy of $\sqrt{s}=13$ TeV.

\section{Conclusions and Outlook}
\label{sec_conclusion}

In this article, we have studied the effects of global conservation laws on the
particle sampling process on the Cooper-Frye hypersurface. This question is
relevant for event-by-event hybrid approaches that are used as the 'standard
model' for the theoretical description of the dynamical evolution of heavy ion
collisions at high beam energies. In addition to the conventional sampling
without conservation of quantum numbers and the existing mode sampling
and SER algorithms, a new SPREW algorithm using exponential weights to ensure global
conservation of quantum numbers has been introduced.

The mode sampling, which is supposed to match the microcanonical ensemble, produces
correct mean multiplicities only in case of the large volume. For a small volume
mode sampling introduces species-dependent biases to means, especially the neutral mesons are affected.
Fluctuations produced by modes sampling are always dramatically larger than the analytical expectation.
The SPREW and SER algorithms correctly match the canonical ensemble, while SER with
rejection by total energy approximately matches the microcanonical expectation. Although
the SER algorithm seems to be the best one in terms of rigour, it is also the slowest one,
especially with the energy rejection. The SPREW algorithm can be a fast alternative
to SER to generate canonical ensemble.

Conserving quantum numbers on the Cooper-Frye transition between hydrodynamics
and hadronic transport is potentially crucial for future studies of particle
correlations and fluctuations both at high and at low beam energies. For example, any higher moment
analysis with hybrid approaches at intermediate or lower beam energies needs to
consider this effect. Therefore, the results of this study might have important
consequences on the interpretation of experimental results related to the QCD
critical point and first order phase transition.

\section{Acknowledgments}
The authors acknowledge funding of a Helmholtz Young Investigator Group
VH-NG-822 from the Helmholtz Association and GSI. This work was supported by the
Helmholtz International Center for the Facility for Antiproton and Ion Research
(HIC for FAIR) within the framework of the Landes-Offensive zur Entwicklung
Wissenschaftlich-Oekonomischer Exzellenz (LOEWE) program launched by the State
of Hesse. Computational resources have been provided by  the GreenIT Cube at
GSI. D. O. acknowledges financial support through a stipend of the Deutsche
Telekom Stiftung.


\begin{thebibliography}{}

\bibitem{Hirano:2012kj}
  T.~Hirano, P.~Huovinen, K.~Murase and Y.~Nara,
  Prog.\ Part.\ Nucl.\ Phys.\  {\bf 70} (2013) 108
  doi:10.1016/j.ppnp.2013.02.002
  [arXiv:1204.5814 [nucl-th]].



\bibitem{Petersen:2014yqa}
  H.~Petersen,
  J.\ Phys.\ G {\bf 41} (2014) no.12,  124005
  doi:10.1088/0954-3899/41/12/124005
  [arXiv:1404.1763 [nucl-th]].



\bibitem{Werner:2010aa}
  K.~Werner, I.~Karpenko, T.~Pierog, M.~Bleicher and K.~Mikhailov,
  Phys.\ Rev.\ C {\bf 82} (2010) 044904
  doi:10.1103/PhysRevC.82.044904
  [arXiv:1004.0805 [nucl-th]].



\bibitem{Shen:2014vra}
  C.~Shen, Z.~Qiu, H.~Song, J.~Bernhard, S.~Bass and U.~Heinz,
  Comput.\ Phys.\ Commun.\  {\bf 199} (2016) 61
  doi:10.1016/j.cpc.2015.08.039
  [arXiv:1409.8164 [nucl-th]].



\bibitem{Ryu:2015vwa}
  S.~Ryu, J.-F.~Paquet, C.~Shen, G.~S.~Denicol, B.~Schenke, S.~Jeon and C.~Gale,
  Phys.\ Rev.\ Lett.\  {\bf 115} (2015) no.13,  132301
  doi:10.1103/PhysRevLett.115.132301
  [arXiv:1502.01675 [nucl-th]].



\bibitem{Pratt:2015zsa}
  S.~Pratt, E.~Sangaline, P.~Sorensen and H.~Wang,
  Phys.\ Rev.\ Lett.\  {\bf 114} (2015) 202301
  doi:10.1103/PhysRevLett.114.202301
  [arXiv:1501.04042 [nucl-th]].



\bibitem{Bernhard:2015hxa}
  J.~E.~Bernhard, P.~W.~Marcy, C.~E.~Coleman-Smith, S.~Huzurbazar, R.~L.~Wolpert and S.~A.~Bass,
  Phys.\ Rev.\ C {\bf 91} (2015) no.5,  054910
  doi:10.1103/PhysRevC.91.054910
  [arXiv:1502.00339 [nucl-th]].



\bibitem{Bernhard:2016tnd}
  J.~E.~Bernhard, J.~S.~Moreland, S.~A.~Bass, J.~Liu and U.~Heinz,
  Phys.\ Rev.\ C {\bf 94} (2016) no.2,  024907
  doi:10.1103/PhysRevC.94.024907
  [arXiv:1605.03954 [nucl-th]].



\bibitem{Oliinychenko:2015lva}
  D.~Oliinychenko and H.~Petersen,
  Phys.\ Rev.\ C {\bf 93} (2016) no.3,  034905
  doi:10.1103/PhysRevC.93.034905
  [arXiv:1508.04378 [nucl-th]].



\bibitem{Keegan:2016cpi}
  L.~Keegan, A.~Kurkela, A.~Mazeliauskas and D.~Teaney,
  JHEP {\bf 1608} (2016) 171
  doi:10.1007/JHEP08(2016)171
  [arXiv:1605.04287 [hep-ph]].



\bibitem{Schenke:2016ksl}
  B.~Schenke and S.~Schlichting,
  Phys.\ Rev.\ C {\bf 94} (2016) no.4,  044907
  doi:10.1103/PhysRevC.94.044907
  [arXiv:1605.07158 [hep-ph]].



\bibitem{Cooper:1974mv}
  F.~Cooper and G.~Frye,
  Phys.\ Rev.\ D {\bf 10} (1974) 186.
  doi:10.1103/PhysRevD.10.186



\bibitem{Ahmad:2016ods}
  S.~Ahmad, H.~Holopainen and P.~Huovinen,
  arXiv:1608.03444 [nucl-th].



\bibitem{Bass:2000ib}
  S.~A.~Bass and A.~Dumitru,
  Phys.\ Rev.\ C {\bf 61} (2000) 064909
  doi:10.1103/PhysRevC.61.064909
  [nucl-th/0001033].



\bibitem{Steinheimer:2017vju}
  J.~Steinheimer, J.~Aichelin, M.~Bleicher and H.~Stöcker,
  arXiv:1703.06638 [nucl-th].

\bibitem{Karpenko:2015xea}
  I.~A.~Karpenko, P.~Huovinen, H.~Petersen and M.~Bleicher,
  Phys.\ Rev.\ C {\bf 91} (2015) no.6,  064901
  doi:10.1103/PhysRevC.91.064901
  [arXiv:1502.01978 [nucl-th]].

\bibitem{Hirano:2005xf}
  T.~Hirano, U.~W.~Heinz, D.~Kharzeev, R.~Lacey and Y.~Nara,
  Phys.\ Lett.\ B {\bf 636} (2006) 299
  doi:10.1016/j.physletb.2006.03.060
  [nucl-th/0511046].



\bibitem{Nonaka:2006yn}
  C.~Nonaka and S.~A.~Bass,
  Phys.\ Rev.\ C {\bf 75} (2007) 014902
  doi:10.1103/PhysRevC.75.014902
  [nucl-th/0607018].



\bibitem{Song:2010mg}
  H.~Song, S.~A.~Bass, U.~Heinz, T.~Hirano and C.~Shen,
  Phys.\ Rev.\ Lett.\  {\bf 106} (2011) 192301
   Erratum: [Phys.\ Rev.\ Lett.\  {\bf 109} (2012) 139904]
  doi:10.1103/PhysRevLett.106.192301, 10.1103/PhysRevLett.109.139904
  [arXiv:1011.2783 [nucl-th]].



\bibitem{Ryu:2012at}
  S.~Ryu, S.~Jeon, C.~Gale, B.~Schenke and C.~Young,
  Nucl.\ Phys.\ A {\bf 904-905} (2013) 389c
  doi:10.1016/j.nuclphysa.2013.02.031
  [arXiv:1210.4588 [hep-ph]].



\bibitem{vanderSchee:2013pia}
  W.~van der Schee, P.~Romatschke and S.~Pratt,
  Phys.\ Rev.\ Lett.\  {\bf 111} (2013) no.22,  222302
  doi:10.1103/PhysRevLett.111.222302
  [arXiv:1307.2539 [nucl-th]].



\bibitem{Pang:2014pxa}
  L.~G.~Pang, G.~Y.~Qin, V.~Roy, X.~N.~Wang and G.~L.~Ma,
  Phys.\ Rev.\ C {\bf 91} (2015) no.4,  044904
  doi:10.1103/PhysRevC.91.044904
  [arXiv:1410.8690 [nucl-th]].




\bibitem{Batyuk:2016qmb}
  P.~Batyuk {\it et al.},
  Phys.\ Rev.\ C {\bf 94} (2016) 044917
  doi:10.1103/PhysRevC.94.044917
  [arXiv:1608.00965 [nucl-th]].
  
\bibitem{Adamczyk:2013dal} 
  L.~Adamczyk {\it et al.} [STAR Collaboration],
  Phys.\ Rev.\ Lett.\  {\bf 112}, 032302 (2014)
  doi:10.1103/PhysRevLett.112.032302
  [arXiv:1309.5681 [nucl-ex]].
  
\bibitem{Aggarwal:2010wy} 
  M.~M.~Aggarwal {\it et al.} [STAR Collaboration],
  Phys.\ Rev.\ Lett.\  {\bf 105}, 022302 (2010)
  doi:10.1103/PhysRevLett.105.022302
  [arXiv:1004.4959 [nucl-ex]].

\bibitem{Steinheimer:2017dpb}
  J.~Steinheimer and V.~Koch,
  arXiv:1705.08538 [nucl-th].

\bibitem{Bozek:2012en}
  P.~Bozek and W.~Broniowski,
  Phys.\ Rev.\ Lett.\  {\bf 109} (2012) 062301
  doi:10.1103/PhysRevLett.109.062301
  [arXiv:1204.3580 [nucl-th]].

\bibitem{Petersen:2008dd}
  H.~Petersen, J.~Steinheimer, G.~Burau, M.~Bleicher and H.~Stocker,
  Phys.\ Rev.\ C {\bf 78} (2008) 044901
  doi:10.1103/PhysRevC.78.044901
  [arXiv:0806.1695 [nucl-th]].



\bibitem{Huovinen:2012is}
  P.~Huovinen and H.~Petersen,
  Eur.\ Phys.\ J.\ A {\bf 48} (2012) 171
  doi:10.1140/epja/i2012-12171-9
  [arXiv:1206.3371 [nucl-th]].

\bibitem{Oliinychenko:2016vkg}
  D.~Oliinychenko and H.~Petersen,
  J.\ Phys.\ G {\bf 44} (2017) no.3,  034001
  doi:10.1088/1361-6471/aa528c
  [arXiv:1609.01087 [nucl-th]].


\bibitem{Pang_inprep}
 L.-G. Pang, in preparation

\bibitem{Hauer:2007ju}
  M.~Hauer, V.~V.~Begun and M.~I.~Gorenstein,
  Eur.\ Phys.\ J.\ C {\bf 58}, 83 (2008)
  doi:10.1140/epjc/s10052-008-0724-1
  [arXiv:0706.3290 [nucl-th]].

\bibitem{Pang:2014ipa}
  L.~G.~Pang, Y.~Hatta, X.~N.~Wang and B.~W.~Xiao,
  Phys.\ Rev.\ D {\bf 91} (2015) no.7,  074027
  doi:10.1103/PhysRevD.91.074027
  [arXiv:1411.7767 [hep-ph]].
  
\bibitem{Huovinen:2009yb}
  P.~Huovinen and P.~Petreczky,
  Nucl.\ Phys.\ A {\bf 837}, 26 (2010)
  doi:10.1016/j.nuclphysa.2010.02.015
  [arXiv:0912.2541 [hep-ph]].


\end{thebibliography}
\end{document}